\documentclass[prd,aps]{revtex4}

\newcommand{\bb}{\begin{eqnarray}}
\newcommand{\ee}{\end{eqnarray}}

\begin{document}

\title{ \bf Relativistic Aharonov--Bohm effect in the presence of two-dimensional
Coulomb potential}
\author{V.R. Khalilov}
\email{khalilov@thc.phys.msu.su}
\affiliation{Physics Department, Moscow State University,
119899, Moscow, Russia}

\begin{abstract}
We obtain exact solutions to the Dirac equation and the relevant
binding energies in the combined Aharonov--Bohm--Coulomb potential
in 2+1 dimensions. By means of solutions obtained the quantum
Aharonov--Bohm (AB) effect is studied for free and bound electron
states. We show that the total scattering amplitude in the the
combined Aharonov--Bohm--Coulomb potential is a sum of the
Aharonov--Bohm and the Coulomb scattering amplitudes. This
modifies expression for the standard Aharonov--Bohm cross section
due to the interference these two amplitudes with each other.
\end{abstract}


\maketitle

\section{Introduction}

The quantum Aharonov--Bohm (AB) effect, predicted by Aharonov and Bohm \cite{1},
was analyzed in many physical
sides in numerous theoretical and experimental works (see,
\cite{2}). The AB effect occurs when electrons travel in a certain
configuration of a vector potential $A_{\mu}$ in which the
corresponding magnetic flux is confined to a finite-radius tube
topologically equivalent to a cylinder. In the case of cylindrical
external field configuration, where a natural assumption is that
the relevant quantum mechanical system is invariant along the
symmetry ($z$) axis, the system then becomes essentially
two-dimensional in the $xy$ plane \cite{khu}.

When an electron travels in an Aharonov--Bohm potential the electron
wave function acquires a (topological) phase which further influences
interference pattern. The vector potential can produce observable effects
because the relative (gauge invariant) phase of the electron wave
function, correlated with a nonvanishing  gauge vector
potential in the domain where the magnetic field vanishes, depends
on only the magnetic flux \cite{khu}. In definite sense one can
say that the Aharonov--Bohm effect is due to the topological
properties of a space of electron wave functions  in a topologically
nontrivial background.

In works \cite{ja,slz} the contribution in the AB amplitude, which
can arise from the inclusion the spin-orbit interaction of the
electron magnetic moment with the electric field oriented along
the solenoid axis, was theoretically studied in the
nonrelativistic approximation. This effect has been recently
confirmed in experiment \cite{jby}. We note that this quantum
system also has the axial symmetry.

Many physical phenomena occurring in quantum systems of
electrically charged fermions, which have the axial symmetry, can
be studied more effectively by means of the corresponding Dirac
equation in 2+1 dimensions that enables one to consider the
relativistic effects.

A permanent interest in this topic also is stimulated by the
studies of (2+1)-dimensional models in both high-temperature
superconductivity \cite{fw} and particle theory (including the
quantum Hall effect \cite{pg} and the degenerate planar
semiconductors with low-energy electron dynamics
\cite{ams})[7-14]).

In section II we study the electron states in an Aharonov--Bohm
potential and discuss the topological properties of a space of
electron wave functions in 2+1 dimensions. In section III  we
obtain exact solutions of the Dirac equation for bound electron
states in 2+1 dimensions in the combined Aharonov--Bohm--Coulomb
potential. In section IV solutions of the Dirac equation for the
scattering problem in 2+1 dimensions in the combined
Aharonov--Bohm--Coulomb potential are obtained. In section V the
Aharonov--Bohm scattering in the combined Aharonov--Bohm--Coulomb
potential are discussed.

We use the units where $c=\hbar=1$.

\section{Electron states in an Aharonov--Bohm potential}

The most wide configuration of external fields, in which exact
solutions of the Dirac equation in 2+1 dimensions are managed to
find in the form of special functions, seems to be the
configuration of an Aharonov--Bohm potential \bb A^0=0,\quad
A_x=-\frac{By}{r^2},\quad A_y=\frac{Bx}{r^2}; \quad A^0=0,\quad
A_r=0,\quad A_{\varphi}=\frac{B}{r},
\quad  B=\frac{\Phi}{2\pi},\nonumber\\
\quad r=\sqrt{x^2+y^2}, \quad
\varphi=\arctan(y/x),\phantom{mmmmmmmmm} \label{eight} \ee and as
well as a vector \bb A^0(r) = -\frac{a}{er}, \quad A_r=0
\label{super} \ee (where $e$ is the electrical charge of a
fermion) and a scalar $U(r)=-b/r$ Coulomb potentials.

In three-dimensional space  such a configuration is the magnetic
field of an infinitely thin solenoid creating a finite magnetic
flux $\Phi$ in the $z$ direction (the magnetic field
$B_z=\Phi\delta({\bf r})$) and the electric field  of a thin
thread oriented along the solenoid axis perpendicular to the plane
$z=0$  and having both electric and scalar charges of constant
linear densities $a/2$ and $b/2$, respectively. The interaction
with a scalar field can be introduced in theory by means of
replacement $m\to m+U$, where $m$ is a fermion mass. In our
problem this scalar interaction is not actual and so below is not
considered.

In 2+1 dimensions, the Dirac $\gamma^{\mu}$-matrix algebra is known to be
represented in terms of the two-dimensional Pauli matrices, $\sigma_j$ \cite{9}.
In addition, two kinds of fermions can be introduced
in accordance with the signature of the two-dimensional Dirac matrices \cite{9}
$$
 \eta = \textstyle\frac{i}{2} {\rm Tr}(\gamma^0\gamma^1\gamma^2) = \pm 1,
$$
where two signs of $\eta$ correspond to two nonequivalent representations
of the Dirac matrices.
We choose
$$
\gamma^0=\eta \sigma_3,\quad \gamma^1=i\sigma_1,\quad \gamma^2=i\sigma_2.
$$
It will be noted that the model with charged fermions is invariant
under the charge conjugation operation and the transformation
$m\to -m$, which is equivalent to the transformation
$\gamma^{\mu}\to -\gamma^{\mu}$ or $\eta\to -\eta$. Hence, we can
fix signs $e$ and $m$.

Let us consider an electron of mass $m>0$ and charge $e$ in the $xy$
plane in potential (\ref{eight}).
The Dirac equation in 2+1 dimensions in potential $A_{\mu}$ is
\bb
(\gamma^{\mu} \hat P_{\mu} - m)\Psi = 0.
\label{two}
\ee
Here $\hat P_{\mu} = -i\partial_{\mu} - eA_{\mu}$ is the generalized electron
momentum operator.

We seek solutions of Eq. (\ref{two}) in (\ref{eight}) in the form
\bb
 \Psi(t,{\bf x}) = \frac{1}{\sqrt{2\pi}}\exp(-iEt+il\varphi)
\psi(r, \varphi)~,
\label{three}
\ee
where $E$ is the electron energy, $l$ is an integer, and
\bb
\psi(r, \varphi) =
\left( \begin{array}{c}
f(r)\\
g(r)e^{i\varphi}
\end{array}\right)~.
\label{four}
\ee
Electron wave function in field (\ref{eight}) (limited at $r=0$) has the form
\bb
 \Psi_p(r, \varphi)=e^{-iEt+il\varphi}\sqrt{\frac{\pi p}{2E}}
\left( \begin{array}{c}
\sqrt{E+\eta m}J_{|\nu|}(pr)\\
-i\sqrt{E-\eta m}e^{i\varphi}J_{|\nu+1|}(pr)
\end{array}\right).
\label{sol1}
\ee
Here $p = \sqrt{E^2 - m^2}$, and $J_{\nu}(pr)$) is the Bessel function
of the index
$$
\nu =|l+eB|.
$$

It should be noted that the irregular solution can be eliminated only at
the definite limitation on admissible values of $|\nu|$. Indeed, in order
that the irregular (the Neumann function $N_{|\nu|}(pr)$) solution can be
eliminated we need to allot it on the ``background'' of the regular solution
$J_{|\nu|}(pr)$ at $r\to 0$ what leads to condition $|l+eB|>0$.

Wave functions are normalized by condition \bb \int
\psi_{p,l,\eta}^*\psi_{p',l',\eta}d^2x=2\pi\delta_{l,l'}\delta(p-p').
\label{norm} \ee At $B=0$ one recover the free electron solutions
in 2+1 dimensions from (\ref{sol1}).

We note that solutions (\ref{sol1}) are one-valued only when the
index $\nu$ is an integer, for example  $l+s$. In this case the
magnetic field flux is quanitized as
$$
\Phi=2\pi \hbar cs/e\equiv \Phi_0 s,
$$
where $\Phi_0$ is the elementary magnetic flux, and $eB=s$.
If $eB$ is not an integer solutions (\ref{sol1}) are many-valued.

One can define the scattering amplitude (SA) in a conventional
manner. We assume that the incident electron wave is from the left and
the wave function is normalized by the standard manner, i.e.
the upper component of the incident wave is $\psi=e^{ipx}$. In fact, the
electron wave function in potential (\ref{eight}) must have at $r\to \infty$
the asymptotic form
\bb
\psi_p(r, \varphi) =
\left( \begin{array}{c}
1\\
-ip/(E+\eta m)
\end{array}\right)e^{ipx+ieB\varphi}
+
\left( \begin{array}{c}
1\\
ip/(E+\eta m)
\end{array}\right)\frac{f(\varphi)}{\sqrt{r}}e^{ipr}.
\label{solscat}
\ee
Here $f(\varphi)$ is the scattering amplitude.

Written $\psi(r, \varphi)$ in the form
\bb
 \psi(r, \varphi)=\sum\limits_{l=-\infty}^{\infty} A_lJ_{|\nu|}(pr)e^{il\varphi},
\label{exsolsc}
\ee
it is easily to show  that we must put
\bb
A_l=e^{-i(\pi/2)|l+eB|}.
\label{ampscat1}
\ee
The scattering amplitude is proportional to $S_l-1\equiv
e^{2i\delta_l}-1$, where $\delta_l=(\nu-l)\pi=eB\pi\equiv
e\Phi/2\hbar c$ are the partial phase shifts. They depend upon
only total magnetic flux $\Phi$.

Coefficient before $e^{ipr}/\sqrt{r}$, is the standard
Aharonov--Bohm amplitude \bb f_{\rm A}(\varphi)=
\frac{1}{\sqrt{2\pi pi}}\frac{e^{-i\varphi(s-1/2)}
\sin(e\Phi/2)}{\sin(\varphi/2)}. \label{ampscat2} \ee Here
$e\Phi=2\pi s+2\pi\Delta, \quad -1/2\le\Delta\le 1/2$.

In order to give ``topological'' properties of solutions one can define
the so-called ``topological'' current
\bb
 J^{\mu}=\psi^* e^{\mu\nu\rho}[-ix_{\nu}\partial_{\rho}+
e\partial_{\nu}A_{\rho}]\psi
= -i\psi^*e^{\mu\nu\rho}x_{\nu}\partial_{\rho}\psi + j^{\mu},
\label{topcur}
\ee
where $\partial_{\mu}\equiv \frac{\partial}{\partial x^{\mu}}$, and
$\psi$ is free electron wave function (at $B=0$).
Currents $J^{\mu}$  and $j^{\mu}$  satisfy continuity equations
$\partial_{\mu}J^{\mu}=\partial_{\mu}j^{\mu}=0$ and, therefore,
\bb Q=\frac{1}{2\pi}\int J^0d{\bf
r}=\int\psi^*\left(-i\frac{\partial}{\partial\varphi}+ 2\pi
eB\delta({\bf r})\right)\psi d{\bf r}, \label{topc1} \ee and \bb
q=\frac{1}{2\pi}\int j^0 d{\bf r}=eB\int\psi^*\delta({\bf r})\psi
d{\bf r} =\frac{eB}{\hbar
c}=\frac{\Phi}{\Phi_0}=\left[\frac{\Phi}{\Phi_0}\right]+
\left(\frac{\Phi}{\Phi_0}\right)_d \label{topc2} \ee (where
$[\Phi/\Phi_0]$ is the integer part of $\Phi/\Phi_0$) are
conserved. Quantity $[\Phi/\Phi_0]$  is the ``topological
number'', and $q-[\Phi/\Phi_0]$ can be called the ``topological
defect''.

Note that the topological quantities introduced here characterize
such properties of solutions as the limit, continuity, and
uniqueness, in difference from the usual topological numbers which are
by the boundary conditions for solutions at $r\to \infty$. The latter
is conserved  due to the finiteness of energy. In our case the ``topological
defect'' is of importance. It occurs due to the branching of solutions in
the point $x, y=0$.
One can as well say that if $eB$ is not an integer than an electron wave
function in potential (\ref{eight}) gains ``topological'' phase.

\section{Bound states}

Now we consider the bound electron states and binding energies in the
combined Coulomb--Aharonov--Bohm potentials.
Looking for solutions in form (\ref{three}) after simple standard
rearrangement, we obtain for $f(r)$ and $g(r)$
\bb
{df\over dr}-{l+eB\over r}f+\left(E+m+\frac{a}{r}
\right)g = 0,
\nonumber \\
{dg\over dr}+{1+l+eB\over r}g-\left(E-m+\frac{a}{r} \right)f = 0.
\label{eq7} \ee
Here and below we put $\eta=1$.
 The functions $f(r)$ and $g(r)$ are normalized
\bb
 \int\limits_0^{\infty}(|f^2|+|g^2|)dr = 1.
\label{eq8}
\ee

Further, following \cite{blp} for $f(r)$ and $g(r)$, we obtain
\bb
 f(r) = \sqrt{m+E}e^{-\rho/2}\rho^{\gamma-1}(Q_1 + Q_2)~, \nonumber \\
 g(r) = -\sqrt{m-E}e^{-\rho/2}\rho^{\gamma-1}(Q_1 - Q_2)~,
\label{sys1}
\ee
where
\bb
 \rho = 2\lambda r,\quad \lambda = \sqrt{m^2-E^2},
\label{notion}
\ee
$\gamma$ is determined by
\bb
\gamma=\frac12\pm\sqrt{\left(l+eB+\frac12\right)^2-a^2},
\label{eq9}
\ee
and solutions  (finite at $\rho=0$) are expressed in terms of the confluent
hypergeometric function $F(b, c; z)$:
\bb
 Q_1 = AF\left(\gamma-\frac12 - \frac{aE}{\lambda}, 2\gamma; \rho\right),
\nonumber\\
Q_2 = CF\left(\gamma+\frac12 - \frac{aE}{\lambda}, 2\gamma; \rho\right).
\label{twfour}
\ee
The constants $A$ and $C$ are related by
\bb
 C =  \frac{\gamma-1/2-Ea/\lambda}{l+eB+1/2+ma/\lambda}A.
\label{conn}
\ee

At  $(l+eB+1/2)^2>a^2$ $\gamma$ is real, and must be chosen positive.
If $a^2>(l+eB+1/2)^2$ then both roots of $\gamma$ are imaginary and
corresponding wave functions oscillate at $r\to 0$.
So the pure Coulomb field in 2+1 dimensions can be considered  only for $(l+eB+1/2)^2>a^2$.

In order to $Q_1$ and $Q_2$ were normalized they must reduce
to polynomials. For $F(b, c; z)$ it means that  $b$ must be equal
to a negative integer or zero, therefore
\bb
 \gamma - \frac12-\frac{Ea}{\lambda} = -n.
\label{spect} \ee
It is easily to show that admitted values of the
quantum number $n$ are: 0, 1, 2,\ldots for $l+eB+1/2>0$ and 1, 2,
3,\ldots for $l+eB+1/2<0$, and the discrete electron energy
levels are given by \bb
 E_n = m\left[1 + \frac{a^2}{(n + \sqrt{(l+eB+1/2)^2-a^2})^2}
\right]^{-1/2}~.
\label{twsix}
\ee

Functions (\ref{twfour}) are one-valued only for integers $eB$.
It is seen from (\ref{twsix}) that the Aharonov--Bohm potential
influences the radiation spectrum of electron.
In addition, if $eB$ are not integers, the phases of electron wave functions
of bound states also depend upon flux parameter $eB$
owing to factor $e^{-iE_n(eB)t}$.

It should be remarked that solutions of Klein--Gordon equation contain parameter
\bb
\gamma^s=\sqrt{(l+eB)^2-a^2},
\label{eq9kg}
\ee
therefore
\bb
 E_n^s = m\left[1 + \frac{a^2}{(n-1/2+\sqrt{(l+eB)^2-a^2})^2}
\right]^{-1/2}~.
\label{twsix1}
\ee
This expression makes sense only when $|l+eB|>a^2$, a condition that
forbids the existence of the $l=0$ energy level at $B=0$.

\section{Aharonov--Bohm scattering in the presence of a Coulomb potential}

The wave functions of the continuous spectrum ($E>m$) can be obtained
from (\ref{sys1}) by means of replacements
\bb
\sqrt{m-E}\to -i\sqrt{E-m},\quad
\lambda\to -ip,\quad -n\to \gamma-1/2-iaE/p.
\label{contin}
\ee
These functions should also be normalized anew.
After replacements (\ref{contin}), let us represent $f$ and $g$ in the form
\bb
\left( \begin{array}{c}
f\\
g
\end{array}\right)
=
\left( \begin{array}{c}
\sqrt{E+m}\\
\sqrt{E-m}
\end{array}\right)
A'e^{ipr}(2pr)^{\gamma-1}[e^{i\xi}F(\gamma-1/2-i\mu,2\gamma,-2ipr)\mp
e^{-i\xi}F(\gamma+1/2-i\mu,2\gamma,-2ipr)],
\label{eqn8}
\ee
where $A'$ is the normalization constant and
$$
\mu=\frac{aE}{p}, \quad e^{-2i\xi}=\frac{\gamma-1/2+i\mu}{l+eB+1/2-i\mu'},
\quad \mu'=\frac{ma}{p}\equiv \mu\frac{m}{E}.
$$
We note that quantity $\xi$ is real.

After simple transformations given for three-dimensional case in \cite{blp},
we obtain
\bb
\left( \begin{array}{c}
f\\
g
\end{array}\right)
=\sqrt{\frac{E\pm m}{Ep}}\frac{(2pr)^{\gamma}}{r}
\frac{|\Gamma(\gamma+1/2+i\mu)|}
{\Gamma(2\gamma)}e^{\pi\mu/2}\phantom{mmmmmmm}\nonumber\\
e^{-i(\pi/2-\gamma+1/2+\mu\ln2pr-\arg\Gamma(\gamma+1/2+i\mu))}
\left( \begin{array}{c}
\Im\\
\Re
\end{array}\right)
[e^{i(pr+\xi)} F(\gamma-1/2-i\mu, 2\gamma, -2ipr)].
\label{eqn6}
\ee
Here $\Gamma(z)$ is $\Gamma$ function.

Asymptotically, the wave function has the form
\bb
\left( \begin{array}{c}
f\\
g
\end{array}\right)
=\sqrt{\frac{2(E\pm m)}{Er}}
\left( \begin{array}{c}
\sin\\
\cos
\end{array}\right)
\left(pr+\delta_l+\mu\ln2pr-\pi l/2\right),
 \label{eqn2} \ee where
\bb \delta_l=\xi-\pi\gamma/2-\arg\Gamma(\gamma+1/2+i\mu)+\pi/4+\pi
l/2, \label{pha1} \ee and \bb
e^{2i\delta_l}=\frac{l+eB+1/2-i\mu'}{\gamma-1/2+i\mu}
\frac{\Gamma(\gamma+1/2-i\mu)}{\Gamma(\gamma+1/2+i\mu)}e^{i\pi(l-\gamma+1/2)}.
\label{pha2} \ee

The expression for the analytical continuation of Eq. (\ref{pha2}) in the
range $E<m$
\bb
e^{2i\delta_l}=\frac{l+eB+1/2+(am)/\lambda}{\gamma-1/2-(aE)/\lambda}
\frac{\Gamma(\gamma+1/2-(aE)/\lambda)}{\Gamma(\gamma+1/2+(aE)/\lambda)}e^{i\pi(l-\gamma+1/2)}
\label{pha3}
\ee
has the poles at the points where
$\gamma+1/2-(aE)/\lambda=1-n, \quad n=1, 2 \ldots$,
as well as at the point $\gamma-1/2-(aE)/\lambda=-n=0$. In these points
the energy levels are discrete. Near the poles with $n\ne 0$, it is easily
to obtain
\bb
e^{2i\delta_l} \approx (-1)^{n+l}\frac{(l+eB+R+1/2)\lambda^3}
{\Gamma(n+1)\Gamma(2\gamma+n)m^2a(E-E_0)}e^{-i\pi(\gamma-1/2)}.
\label{ph4}
\ee

The residue of function $\exp(2i\delta_l)$
in its pole is related to the coefficient in the asymptotic expression
of the wave function  of the corresponding bound state as follows
\bb
f\approx A_0e^{-\lambda r},\quad g=\sqrt{\frac{m-E}{m+E}}f.
\label{ph1}
\ee
where
\bb
A_0 = \left[\sqrt{\frac{m+E}{m-E}}\frac{(l+eB+ma/\lambda+1/2)\lambda^3}
{2m^2a\Gamma(n+1)\Gamma(2\gamma+n)}\right]^{1/2}
(2\lambda r)^{\gamma+n-1/2}.
\label{ph0}
\ee

Now we consider the scattering problem in 2+1 dimensions in the combined Coulomb--Aharonov--Bohm
potentials. The total phase shifts according to (\ref{eqn2}) are
\bb
\delta_l=-\pi\gamma/2+\pi/4+\pi l/2+\xi-\arg\Gamma(\gamma+1/2+i\mu)\equiv
\delta_{\rm AB} + \delta_l^a,
\label{separ}
\ee
where
\bb
\delta_{\rm AB}=-\pi\gamma/2+\pi/4+\pi l/2
\label{separ1}
\ee
and
\bb
\delta_l^a=\xi-\arg\Gamma(\gamma+1/2+i\mu)
\label{separ2}
\ee
are the phase shifts due to the Aharonov--Bohm  and the Coulomb potential,
respectively. Note that $\delta_{\rm AB}$ and $\delta_l^a$ weakly depend
on $a$ and $eB$, respectively.

The total scattering amplitude is proportional to
\bb
f_{tot}(\varphi)\sim\sum\limits_{l=0}^{\infty}[\exp(2i\delta_{\rm A}+2i\delta_l^a)-1].
\label{amptot}
\ee
The difference in the square brackets is written in the form \cite{ll}
\bb
\exp(2i\delta_{\rm AB}+2i\delta_l^a)-1= [\exp(2i\delta_{\rm A})-1]+
[\exp(2i\delta_{\rm AB})(\exp(2i\delta_l^a)-1)].
\label{amptot1}
\ee

The Coulomb phases mainly contribute in the scattering amplitude
at large $l$, so their contribution can be calculated in the
quasi-classical approximation. After simple calculations and
taking into account Eq.(\ref{ampscat2}), we obtain \bb
f_{tot}(\varphi) =\frac{1}{\sqrt{2\pi
pi}}\frac{1}{\sin(\varphi/2)} \left[\sin(\pi
eB)e^{-i\varphi(s-1/2)}+\frac{am}{p}e^{i\pi eB}\right] \equiv
f_{\rm AB}(\varphi)+f_a(\varphi). \label{amptot2} \ee

From Eq. (\ref{amptot2}) it follows that these two amplitudes interference
with each other in the scattering cross section:
\bb
d\sigma =|f_{tot}(\varphi)|^2d\varphi= \frac{d\varphi}{2\pi p\sin^2\varphi/2}
\left[\sin^2\pi eB
+ \left(\frac{2am}{p}\right)\sin\pi eB\cos(s\varphi+\varphi/2+\pi eB) +
\left(\frac{am}{p}\right)^2\right].
\label{scatter}
\ee
From Eq. (\ref{scatter}) it is seen that the periodic dependence
of the interference term in the cross section differs for toward
($\varphi=0$) and backward ($\varphi=\pi$) scattering.

\section{Resume}

It is shown that  the gauge-invariant (observable)  quantities are
the quantum-mechanical phases of electron wave functions and the
energies of bound states in both  quantum and classical mechanics.
The gauge-invariant phase is the phase,
which acquires the electron wave function when the electron travels
along a closed path which encircles a thin solenoid (oriented along the axis $z$) in
the plane $z=0$.  It will be recalled
that the quantum wave associated with each electron in the
entrance splits into two wave packets that go around the solenoid
with different sides. The ways of these wave packets intersect in
the exit to result in a closed contour.  So, though the
Aharonov--Bohm potential satisfies equation $[\mathbf{\nabla\times
A}]=0$ everywhere in the plane  except the point $x=y=0$, the
integral that gives the magnetic flux $\Phi$ through a closed
contour $C$ enclosed by the wave packets
$$
\oint {\bf A}d{\bf s}=\Phi
$$
is defined unambiguously.

Classical electron (with the energy $E$ and the moment of momentum $L_0$ with respect to axis
$z$) trajectory in the Aharonov--Bohm potential is line
$$
r=\frac{r_{min}}{\cos(\varphi-\varphi_0)},\quad
r_{min}=\frac{L_0c+eB}{\sqrt{E^2-m^2c^4}}
$$
and the scattering angle is zero.

Electron energy expressed via the action variables $J_r$ and $J_{\varphi}$ is
$$
 E_n = mc^2\left[1 + \frac{a^2}{(cJ_r + \sqrt{(cJ_{\varphi}+eB)^2-a^2})^2}
\right]^{-1/2}~.
$$
After quantization ($J_r=\hbar n, \quad J_{\varphi}=\hbar(l+1/2)$) it follows
from this formula Eq.(\ref{twsix}).


\vspace{1cm}

\begin{thebibliography}{55}

\bibitem{1} Y. Aharonov and D. Bohm, Phys. Rev., {\bf 115}, 485 (1959).

\bibitem{2} For a review see, M. Peshkin and A. Tonomura,
{\sl The Aharonov-Bohm Effect}, (Springer-Verlag, Berlin, 1989).

\bibitem{khu} K. Huang, {\sl Quarks, Leptons, and Gauge Fields}, (World
Scientific, Singapore, 1982).

\bibitem{slz} S-L. Zhu and Z.D. Wang, Phys. Rev. Lett., {\bf 85}, 1076 (2000).

\bibitem{ja} J. Anandan, e-print archive: quant-ph/0212102 v1.

\bibitem{jby} Jeng-Bang Yau, E.P. De Poortere, and M.Shayegan,
Phys. Rev. Lett., {\bf 88}, 146801 (2002).

\bibitem{fw} F. Wilczek, {\sl Fractional Statistics and Anyon
Superconductivity}, (World Scientific, Singapore, 1990).

\bibitem{pg} R.E. Prange and S.M. Girvin, eds., {\sl The Quantum
Hall Effect} (2nd ed.), (Springer-Verlag, New York, 1990).

\bibitem{ams} A.M.J. Schakel and G.W. Semenoff, Phys. Rev. Lett.,
{\bf 66}, 2653 (1991).

\bibitem{djt} S. Deser, R. Jackiw, and S. Templeton, Ann. Phys., (NY),
{\bf 140}, 372 (1982).

\bibitem{3} A.J. Niemi and G.W. Semenoff. Phys. Rep. {\bf C135}, 99 (1986).

\bibitem{9} Y. Hosotani, Phys. Lett., {\bf B319}, 332 (1993).

\bibitem{8} V.R. Khalilov and C.L. Ho, Mod. Phys. Lett., {\bf A13}, 615 (1999).

\bibitem{11} C.L. Ho and V.R. Khalilov, Phys. Rev., {\bf A61}, 032104 (2000).

\bibitem{blp} V.B. Berestetzkii, E.M. Lifshitz, and L.P. Pitaevskii,
{\sl Quantum Electrodynamics}, 2nd ed. (Pergamon, New York, 1982).

\bibitem{ll} L.D. Landay, E.M. Lifshitz.
{\sl Quantum Mechanics}, 2nd ed. (Pergamon, New York, 1978)

\end{thebibliography}
\end{document}